# Localized Mobile Agent Framework for data processing on Internet of Things

J. Mahalakshmi
Department of Computer Science
Bharathiar University
Tamilnadu, India
Mahalakshmi1203@gmail.com

P. Venkata Krishna
*Senior Member IEEE*
Department of Computing Science
Sri Padmavati Mahila Visvavidyalayam
Tirupati, India
dr.krishna@ieee.org

Abstract: Internet of Things (IoT) is the major research filed in the recent trends. IoT has the ability to create communication with any object. IoT produces big amount of raw data at the time of data gathering. Therefore, there is a need of efficient mechanism to address the issue of IoT. This paper presents the localized MapReduce Framework for IoT. This frame work processes the data at the local nodes without transferring the data to the cloud or high end servers. The mobile agents have the ability to migrate from one node to another node for data processing, Mapper and Reducer are used for agent duplication and result aggregation. The performance of the proposed framework is evaluated using the cost function and the results proved the efficiency the proposed framework.

1. Introduction

In the recent trends, Internet of Things (IoT) [1] is ranked as primary research field. The IoT is a paradigm which connects different objects, observes their behaviour and collects the sensed data. IoT collects zetta bytes of raw data from the sensor nodes. Therefore, data analysis is a very important aspect that has to be consideringfor efficient IoT management. Big data had the efficient mechanism to fulfil the data needs of IoT. Big data [2] collects and process huge volumes of data which is generated by the sensor nodes in the IoT. The network connectivity in the IoT is operated with low bandwidth; therefore the data transfer through this network should be minimal [13-15].

In IoT, data processing is treated as the complex task, many researches contributed number of computation methods and some models are good enough to solve the issue as much as possible. To address the data processing [16], Google developed MapReduce model which is developed with an idea of map and reduce functions in the LISP programming language. Yahoo developed Hadoop platform which actually uses the MapReduce function for processing the large volumes of datasets [3-4]. The MapReduce framework duplicates the data to the high end servers or to the cloud platform before the data processing is initialized. This process requires huge data storage and it is more cost effectively. The data transmission from the sensor nodes to the high end server requires more cost [5].

To solve this issue, we developed a localized MapReduce framework which is processed at the sensor nodes in IoT environment. The main gaol of the proposed localized framework is to implement the MapReduce at the sensor nodes rather than at the high end servers or cloud and aggregate the findings.

The rest of the paper is organized as follows. Section 2 deals with related work regarding the frameworks presented in the recent years. Section 3 deals with requirements of the framework which has to be achieved at the time of designing. Section 4 explains about the proposed architecture with master and slave policy. Section 5 evaluates the performance of the proposed framework and finally the conclusion is drawn in section 6.

2. Related Work

The development of big data made many opportunities to the business environment for solving the data needs, but the MapReduce [5] framework introduced new approach for data processing using map and reduce process. The MapReduce process divides the data into smaller parts and assigns them to the worker nodes, the worker nodes computes the problem. The results are collected from the worker nodes and aggregate them as the solution to the problem [17-18].

Many commercial service providers are using the MapReduce process by implementing their own file system such as Hadoop file system (HDFS), shared memory from the CPUs and Google File System (GFS). For instance, Hadoop platform uploads the data from the current environment to the HDFS file system before processing it. This process requires more bandwidth and consumes high cost. In our approach, the data processing is done at the local nodes by deploying the code for processing the data with the utilization of localized data. In [6], the author presented the Hadoop framework for embedded computers. The Hadoop file system is not suitable for the high end servers. Therefore, the model doesn't serve good.

Many researchers contributed towards the development of MapReduce framework. For instance, MATE[7] and Phoenix[8] are the models which support the shared memory for CPUs. Some of the contributions are concentrated on iterative execution of MapReduce function, such as MapReduce with access patterns (MRAP) [9], Twister [10] and Haloop [4]. These models store the data in temporary files instead of using the key value storage (KVS).In [11], the Misco framework was developed for data processing in mobile devices using HTTP. The authors in [12] proposed a MapReduce framework for heterogeneous devices. This model concentrated on computing the data at mobile devices.

3. Requirements for the framework

In Hadoop environment, the MapReduce process is implemented in the high end servers, but the IoT environment is completely different from Hadoop environment, The IoTenvironment consists of sensor nodes, the data collection processing and aggregation are done at the sensor nodes. The major requirements of the proposed framework are given as follows:

- ➢ Sensing units of IoT: The sensor nodes in the IoT environment are composed of less computation capacity with low memory when compared to the high end servers.
- ➢ Limited bandwidth: The nodes in the IoT network are operated in low bandwidth. Sometimes, it leads to network failures. Our proposed framework should be tackle with this situation.
- ➢ Each sensor node must have the ability to process the tasks.
- ➢ The IoT network doesn't support the file system. Therefore, the data should be processed at the local storage of nodes.
4. Framework for localized MapReduce in IoT

The proposed framework is completely different from the existing MapReduce Framework. The major advantages of proposed framework are explained as follows:

- ➢ Parallel Execution: In the proposed model, the data should be processed parallel on all nodes and it is assumed that all the nodes are autonomous and execute the data separately.
- ➢ Data processing at senor nodes: In IoT, the sensor nodes are responsible to generate and process the data. Therefore, there is no need to transfer the data to the high end servers or cloud

The proposed framework assumes the nodes are independent and provides parallel execution. Therefore, there is no need of data exchange between the nodes.

4.1. Mobile Agents

In the proposed framework, mobile agents are introduced to achieve the requirements presented in the previous section. The functionality of mobile agents is to carry the tasks and process the tasks at sensor nodes and finally to obtain the results from the sensor nodes. In the original MapReduce model, we have mapper and reducer classes to support the data processing task. In the same way, the proposed framework also has the mapper and reducer. The mobile agents have many advantages compared to the previous ones. The mobile agents can select one or more nodes according to the result and performs migration. This type of practice is very useful in unstable network. The mobile agent does not follow the centralized approach and they operated in decentralized manner. This is crucial to process the data stored in the sensor nodes to maintain scalability. The mobile agents can store and retrieve the data from the heap memory of the senor nodes. Therefore, it is easy for the developers to use the data without knowing the file system like Hadoop.

4.2. Master-Slave Architecture for Proposed framework

The tradition MapReduce framework consists of one master agent and many more slave agents. In the proposed framework, mapper, reducer and slave nodes are deployed at the sensor nodes which are called as mapper nodes, reducer nodes and slave nodes. The mobile agents can dynamically deploy at the senor nodes to process the data based on the available resources. Figure 1 explains about the Mater-Slave architecture of proposed framework.

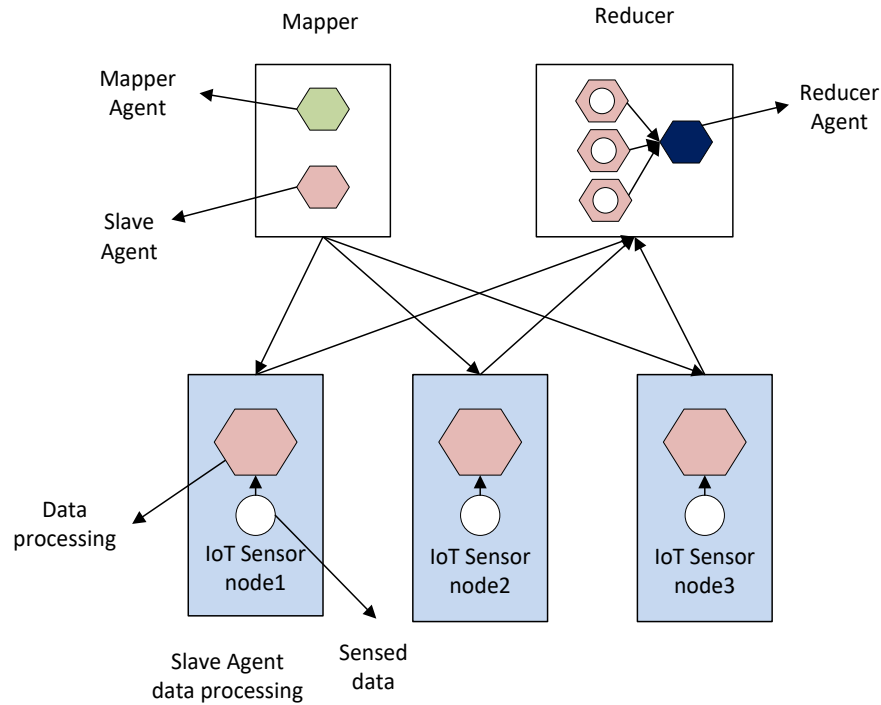

Figure 1: Master –Slave Architecture for proposed Framework

4.3 Algorithm for data processing

The proposed framework is modelled with the enhancement of mobile agents to MapReduce process. Algorithm1 shows how the data processing is carried in the proposed framework.

Algorithm 1: Data processing

Input: Mapper Agent, Slave Agent

Output: aggregated resultant data

Begin

Step1: The master node contains both mapper agent and slave agent, the mapper agent makes copies of slave agents.

Step 2: The slave agents migrates from master node to one or more sensor nodes where the data has to be processed.

Step 3: The slave agents computes the data gathered by the sensor nodes.

Step 4: After the data computation is completed, the slave agents migrates to the reducer node where the final result has to be aggregates.

Step 5: The slave agents utilizes the inter agent communication for sending the results to the reducer.

End

The resultant data obtained from the slave agents are much less than the gathered data at sensor nodes, each slave agent is executing independently and Mapper and reducer are executing on the same Master node.

The major advancement in the proposed framework is utilization of mobile agents. The mobile agents had the ability to migrate between the nodes with their own knowledge. The mobile agents identify the next node after completion of the data processing module independently. For instance, slave agent is allocated to sensor node1, the agent doesn't find any data there to process, and the slave node automatically migrates to the other node. The mapper agent and agent are also act as slave agent i.e, they also migrate from one node to another node. Mapper and Reducer have their own set of program to define the data inside them.

5. Performance Evaluation

The proposed framework is developed by using the java development kit 1.8. The graphical user interface was developed for managing the mobile agents. The performance of the framework is evaluated by using the agent duplication cost function. Here, we are considered 8 nodes with dual core processors along with Giga Ethernet.

Figure 2 shows the comparison of agent size and the cost for agent duplication which is measured in milliseconds. Here, only two callback methods are invoked for agent migration.

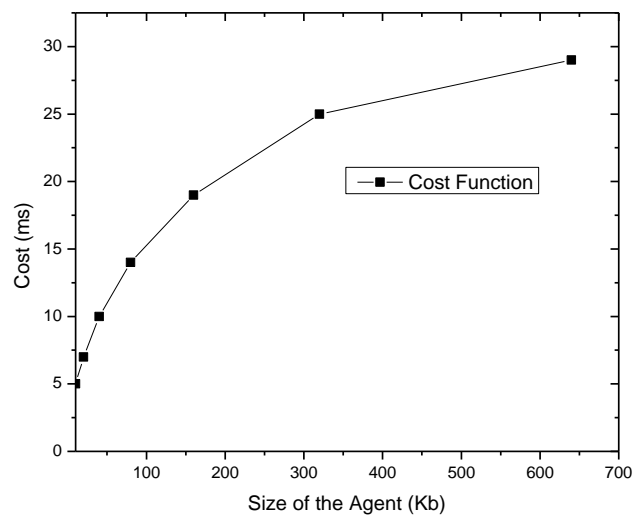

Figure 2: Cost function for Agent duplication

Figure 3 shows the migration cost of agent between two nodes. The migration cost involves the process of preparing the agents, opening the TCP connection, migrating the agents from one node to another node,unpack the agents and security verification. Therefore the cost function is high for agent migration.

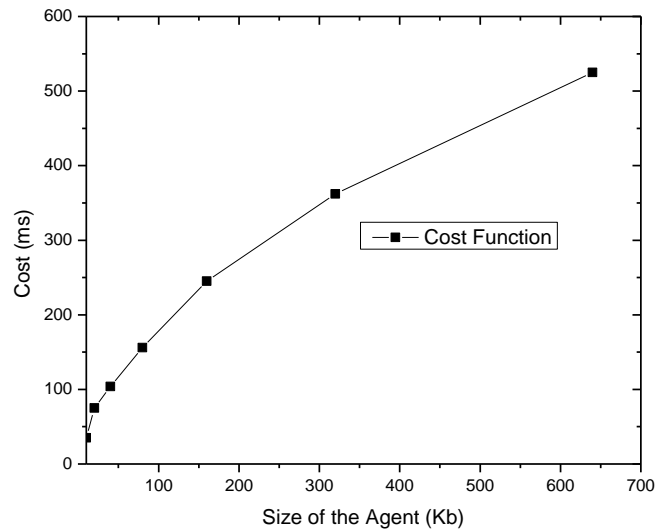

Figure 3: Cost Function for Agent migration

We compared the performance of proposed framework with Hadoop model. The Hadoop model is tested with dual core processor along with 2 GB of RAM and it copies the data with 34MB/s. the amount of data need to transferred is 1.8 GB. The experiment is evaluated with 8 nodes and the total cost recorded for data transmission is 105 seconds from sensor nodes to the HDFS. The total cost for data processing is recorded as 158 seconds. In the proposed framework, the data need to be transferred from nodes are reduced to 16.4MB. This is due to the data processing at the nodes. The proposed model is more efficient for IoT to avoid unnecessary data by pre-processing.

6. Conclusion

This paper proposed the localized MapReduce framework for IoT. The proposed model is unique in their nature because of performing the data processing at the nodes instead of transferring the data to the cloud or high end servers. The traditional MapReduce model is enhanced with the mobile agents. The mobile agents have the ability to migrate from one node to another node for data processing, Mapper and Reducer are used for agent duplication and result aggregation. The proposed framework is useful for the developers to define application with any knowledge of mobile agents. In the future, this model is enhanced with scheduling mechanisms for distributing the tasks for sensor nodes.